\documentclass[aip,jcp, twocolumn,amsmath,amssymb,reprint]{revtex4}
\usepackage{graphicx}
\usepackage[squaren]{SIunits}

\DeclareMathOperator{\tr}{Tr}  

\newcommand{\bea}{\begin{eqnarray}}
\newcommand{\eea}{\end{eqnarray}}

\newcommand{\Eq}[1]{Eq.~\eqref{#1}}

\newcommand{\be}{\begin{equation}}
\newcommand{\ee}{\end{equation}}
\newcommand{\transpose}{^{\text{T}}}
\newcommand{\rcorr}{r_{\text{corr}}}
\newcommand{\rcutoff}{r_{\text{cutoff}}}
\def\<{\left\langle}                 
\def\>{\right\rangle}                 

\begin{document}
\title{A fast Variational Gaussian Wave-packet method: Size-induced structural transitions in large neon clusters.}
\author{Ionu\c{t} Georgescu}
\author{Vladimir A. Mandelshtam}
\affiliation{Chemistry Department,
University of California, Irvine, CA 92697, USA}
\begin{abstract}
The Variational Gaussian wavepacket (VGW) method is an alternative to Path Integral Monte-Carlo (PIMC) for the computation of thermodynamic properties of many-body systems at thermal equilibrium. It provides a direct access to the thermal density matrix and is particularly efficient for Monte-Carlo approaches, as for an $N$-body system it operates in a non-inflated $3N$ dimensional configuration space. Here we greatly accelerate the VGW method by retaining only the relevant short-range correlations in the (otherwise full) $3N\times 3N$ Gaussian width matrix without sacrificing the accuracy of the fully-coupled VGW method. This results in the reduction of the  original $\mathcal{O}(N^3)$ scaling to $\mathcal{O}(N^2)$. The Fast-VGW method  is then applied to quantum Lennard-Jones clusters with sizes up to $N=6500$ atoms. Following Doye and Calvo [JCP {\bf 116}, 8307 (2002)] we study the competition between the icosahedral and decahedral structural motifs in Ne$_N$ clusters as a function of $N$.
\end{abstract}
\maketitle

\section*{Introduction}
\label{introduction}

The Variational Gaussian wavepacket (VGW) method\cite{frantsuzov2003,frantsuzov2004} was introduced recently as an alternative to Path Integral techniques for the estimation of thermodynamic and structural properties of large many-body systems at thermal equilibrium. The VGW method provides a direct and numerically efficient estimate of the density matrix $e^{-\beta\hat{H}}$, and particularly its diagonal elements $\rho(x)=\langle x|e^{-\beta\hat{H}}|x\rangle$. As such it has been combined successfully with Monte-Carlo techniques for computations of thermodynamic and structural properties of atomic and molecular clusters\cite{frantsuzov2006,deckman2008,frantsuzov2008,deckman2009,deckman2009a,deckman2010}. When compared with accurate Path-Integral Monte-Carlo (PIMC) calculations for certain neon clusters, the VGW method yielded practically identical results \cite{frantsuzov2004,predescu2005}. The Thermal Gaussian Molecular Dynamics\cite{georgescu2010,georgescu2011} (TGMD) was built on top of the VGW for the estimation of time correlation functions in quantum many-body systems. Other quantum dynamics approaches, such as the Full Wigner dynamics \cite{liu2007},  Equilibrium Liouville Dynamics \cite{liu2011a,liu2011b} also take advantage of the analytically convenient representation of the density matrix in the VGW formalism.

Given an $N$-body system, the most accurate version of the VGW method, the so-called Fully-Coupled VGW (FC-VGW) utilizes the full $3N\times 3N$ Gaussian width matrix $G$, with the matrix-matrix multiplication step being the numerical bottleneck. The $\mathcal{O}(N^3)$ numerical scaling of the latter limits the applicability of the FC-VGW to systems of relatively small size. The so-called Single-Particle VGW (SP-VGW) utilizes the block-diagonal form of $G$, each particle represented by a $3\times 3$ block. Consequently, the matrix-matrix multiplication is no longer a numerical bottleneck in the SP-VGW, and for systems with short-range pair potentials the overall scaling of the SP-VGW can be reduced to linear in $N$ (after appropriate potential cut-offs are implemented). The drawback of the SP-VGW may be an uncontrollable loss of accuracy (compared to the FC-VGW).
 As shown below, only relatively short-range correlations in the $G$-matrix are significant, even for systems with quite pronounced quantum character, such as (\emph{para}-H$_2$)$_{N}$ clusters. Here we exploit this circumstance and present an improved version of the VGW method, called Fast-VGW, which reduces the overall numerical scaling of the FC-VGW from $\mathcal{O}(N^3)$ to at least $\mathcal{O}(N^2)$ by retaining only the physically relevant off-diagonal elements of the (otherwise full) $G$-matrix, with practically no accuracy tradeoff. (Depending on the system, further reduction in the numerical scaling can be achieved by applying additional cut-offs and utilizing the sparsity of the Hessian matrix.)

In the next section the VGW formalism is introduced. We then analyze the accuracy of the Fast-VGW approach by applying it to the ground state calculations of several Lennard-Jones (LJ) clusters for a range of quantum characters, from moderate (Ne) to relatively strong (\emph{para}-H$_2$). The last section presents an application to very large Ne$_N$ clusters with up to $N\sim 6500$ atoms, where the competition between the icosahedral and decahedral structural motifs for the ground state is studied as a function of the cluster size $N$. In particular, this study provides an improved estimate of the icosahedral-decahedral crossover size for Ne$_N$ clusters compared to that by Calvo and Doye \cite{doye2002}, who used the Harmonic-Superposition Method (HSM) to study the structural and thermodynamic properties of quantum LJ clusters for a range of quantum parameters.

\section*{The Variational Gaussian Wave-packet Approximation}
\label{vgw:formalism}
The matrix elements of the density matrix
\be
\rho(x, x'; \beta) := \< x | e^{-\beta\hat{H}} | x' \> = \< x;\beta/2 | x';\beta/2\>
\ee
can be expressed in terms of the wave-packets $|x,\tau\rangle$
\be
|x;\tau\rangle = e^{-\tau\hat{H}}|x\rangle,
\ee
which are solutions of the imaginary-time Schr\"{o}dinger equation (or Bloch equation):
\be
-\frac{\partial}{\partial \tau} |x;\tau\rangle = \hat{H} |x;\tau\rangle.
\ee

The VGW approach approximates these wavepackets with Gaussians
\bea\label{eq:Gauss}
\langle  r| x;\tau\rangle &\approx&
  {(2\pi)^{-3N/2}\| G\|^{-1/2}} \\\nonumber &\times& \exp\left[ -\frac
    1 2  
 (r-q)^{\rm T}G^{-1} (r-q) +\gamma\right],
\eea
where the Gaussian parameters are $ G= G(\tau)\in \mathbb{R}^{3N\times 3N}$,  
the real symmetric and non-negative Gaussian width matrix, $ q= q(\tau)\in \mathbb{R}^{3N}$,   
the Gaussian center, and $\gamma=\gamma(\tau)\in \mathbb{R}$, the scale factor. 

With this ansatz the diagonal element of the density matrix, or just density $\rho( x;\beta):=\rho(x, x; \beta)$, becomes a simple expression in terms of $G$ and $\gamma$:
\be
\label{eq:K1}
\rho( x;\beta)  = \frac{e^{2\gamma(\beta/2)}}{
  (4\pi)^{3N/2}\| G(\beta/2)\|^{1/2}} \ .
\ee

A variational principle\cite{frantsuzov2004,predescu2005} provides the equations of motion for the Gaussian parameters $q(\tau)$, $G(\tau)$ and, respectively, $\gamma(\tau)$
\begin{subequations}
\begin{align}
\frac {d}{d\tau}  q &= - G\, \langle \nabla  U \rangle  \label{vgw:q}\\
\frac {d}{d\tau}  G &= - G\, \langle \nabla\nabla\transpose U \rangle G + \hbar^2 M^{-1} \label{vgw:g} \\
\frac {d}{d\tau} \gamma &= - \frac{1}{4} \tr\left[\langle \nabla\nabla\transpose U \rangle  G\right] - \langle U \rangle \label{vgw:gamma}
\end{align}
\end{subequations}
which are propagated  with the initial conditions
\be\label{eq:init}
 q(0) =  x,\quad G(0) = 0,\quad \gamma(0)=0
\ee
from $\tau=0$ to $\tau=\beta/2$. By $\langle U \rangle$, $\langle \nabla  U \rangle$ and $\langle \nabla\nabla\transpose U \rangle$ we defined, respectively, the expectation values of the total energy, its gradient and its Hessian, e.g.
\be  
\langle \hat{U} \rangle = \frac{\langle x;\beta/2|\hat{U}|x;\beta/2\rangle} {\langle x;\beta/2|x;\beta/2\rangle}.
\ee

In the zero temperature limit $\beta\rightarrow\infty$, the VGW minimizes the functional
\be
E = \frac{\langle x;\beta/2|\hat{H}|x;\beta/2\rangle} {\langle x;\beta/2|x;\beta/2\rangle}=-\frac{\partial}{\partial\beta}\mbox{ln}\rho(x;\beta)
\label{gse}
\ee
and thus provides a way for estimating both the ground state energy and the density.

The VGW is exact in the harmonic and  classical (high temperature) limit; the analytical solution of the multi-dimensional harmonic oscillator is provided in Ref. \cite{georgescu2011}. Moreover, if $U$ consists of only pair interactions and if these can be fitted in terms of
Gaussians, then $\langle U \rangle$, $\langle \nabla  U \rangle$ and,
respectively, $\langle \nabla\nabla\transpose U \rangle$ can be expressed analytically and as such
are easy to compute. Fortunately, most of the pair potentials used in practice, such as the LJ or Coulomb, or the Silvera-Goldman \cite{silvera1978} potentials can be fitted very
accurately using a small number of Gaussians.

The quantum character, better said, the degree of quantum delocalization of a system is conveniently described by the de Boer parameter
\begin{equation}
\Lambda=\frac{\hbar}{\sigma\sqrt{m\epsilon}},
\end{equation}
which relates the quantum delocalization of the wave-functions, estimated by the de Broglie wave-length $\hbar/\sqrt{m\epsilon}$, to the characteristic length $\sigma$. The latter characterizes the range and $\epsilon$, the strength of the pair potential, for example the LJ potential is
\begin{equation}
U(r_{ij}) = 4\epsilon\left[\left(\frac{\sigma}{r_{ij}}\right)^{12} - \left(\frac{\sigma}{r_{ij}}\right)^{6} \right].
\end{equation}
The classical limit is obtained for $\Lambda \ll 1$. Typical values are $\Lambda=0.01$ for Xe, $\Lambda=0.03$ for Ar, $\Lambda=0.095$ for Ne and $\Lambda\sim 0.3$ for \emph{p}-H$_2$. Rewriting the VGW equations (\ref{vgw:q}-\ref{vgw:gamma}) in reduced units leads to a general set of equations with only one free parameter, $\Lambda$:
\begin{subequations}
\begin{align}
\frac {d}{d\tau}  q &= - G\, \langle \nabla  U \rangle  \label{vgw:qp}\\
\frac {d}{d\tau}  G &= - G\, \langle \nabla\nabla\transpose U \rangle G + \Lambda^2 \label{vgw:gp} \\
\frac {d}{d\tau} \gamma &= - \frac{1}{4} \tr\left[\langle \nabla\nabla\transpose U \rangle  G\right] - \langle U \rangle \label{vgw:gammap}
\end{align}
\end{subequations}
This set of equations can be used to study a wide range of species by varying just one parameter, $\Lambda$.

\section*{The Fast-VGW method: Numerical tests.}

In the most favorable case depicted so far, namely when $U$ consists of pair interactions that can be fitted with Gaussians, the computation of the expectation values of the energy, its gradient and, respectively, its Hessian require $\mathcal{O}(N^2)$ operations. Various algorithms, such as tree-codes \cite{barnes1986}, can reduce this effort to $\mathcal{O}(N\log N)$ or even $\mathcal{O}(N)$ \cite{greengard1987}. With short-range potentials, such as LJ or Silvera-Goldman, the $\mathcal{O}(N)$ scaling can be achieved easily with a potential cut-off.

\begin{figure}[htbp]
	\begin{centering}
		\includegraphics[width=0.8\linewidth]{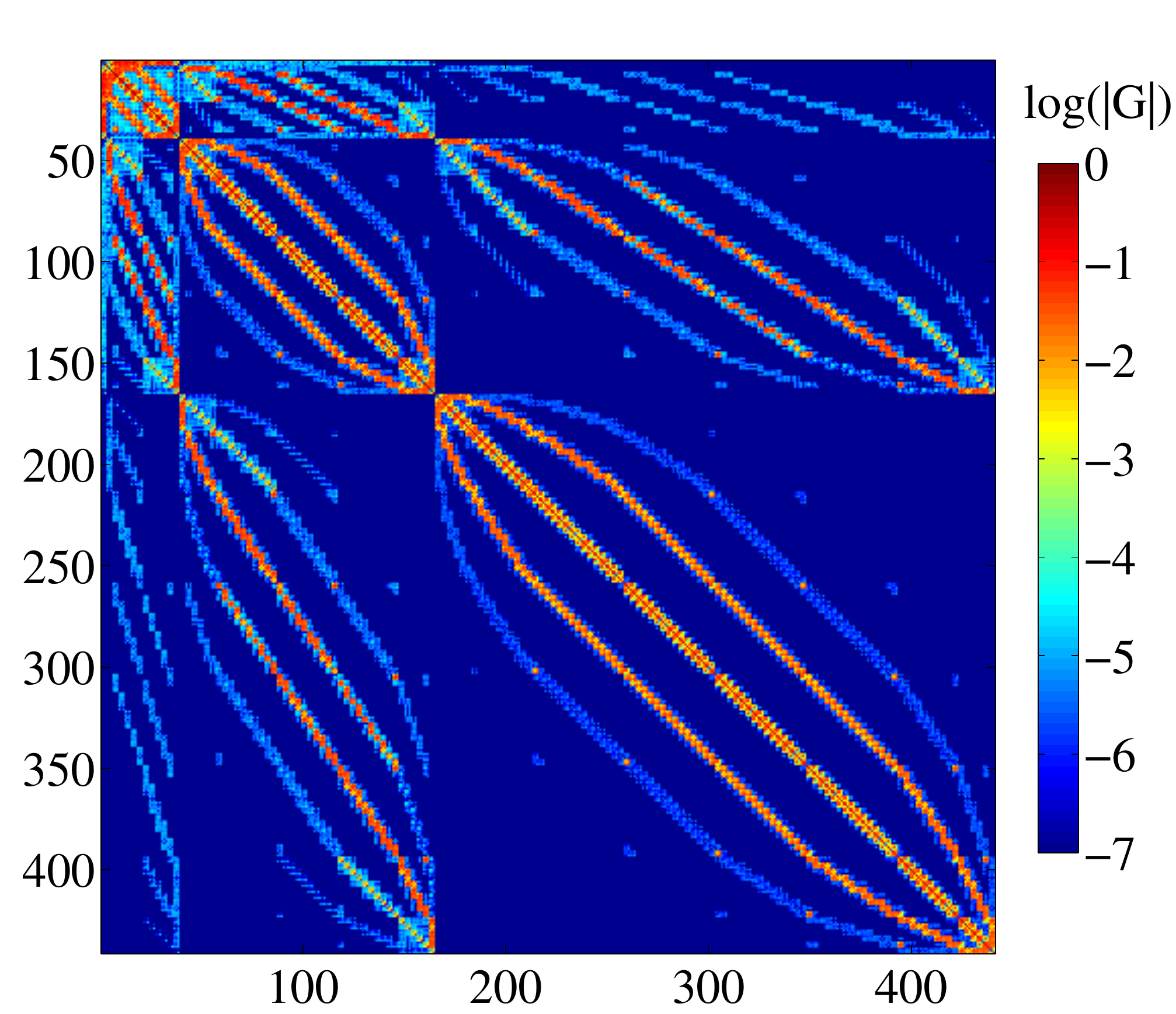}
	\end{centering}
	\caption{The $G$ matrix for Ne$_{147}$ at a temperature of $T=3.56K$ ($T=0.1\epsilon$) within the full, $3N\times 3N$, Gaussian approach (FC-VGW). The color density is proportional to the decimal logarithm of the matrix elements.\label{G147}}
\end{figure}

Yet, the dominant operation in Eqs.~(\ref{vgw:qp}-\ref{vgw:gammap}) is the matrix-matrix multiplication in  Eq.~\eqref{vgw:gp} which scales as $\mathcal{O}(N^3)$, assuming both the  Hessian and the $G$ matrix being general, dense matrices. This sets a practical limit on the size of the system to be treated numerically using the VGW method. (Note also that \Eq{eq:K1} requires calculation of the determinant of $G$, which also scales as $\mathcal{O}(N^3)$ for a full matrix, but this calculation is only performed once in order to compute the density at $\tau=\beta/2$.) For sufficiently small values of the quantum parameter $\Lambda$,  the fully-coupled Gaussian wave-packet can be approximated with product of single-particle Gaussians, thus reducing $G$ to a $3\times 3$ block diagonal form. Refs. \cite{frantsuzov2004,predescu2005} give some idea on how the SP-VGW performs for Ne ($\Lambda\sim 0.1$) clusters, by comparing it to both the FC-VGW and PIMC results.  While the SP-VGW seems to correctly describe the thermodynamics of Ne clusters, it may actually fail, even for such weakly quantum systems, to adequately characterize different cluster configurations with very close energy values. For example, this happens for the Ne$_{38}$ cluster, for which the FC-VGW is still very accurate \cite{predescu2005}.

\begin{figure}[htbp]
\begin{centering}
\includegraphics[width=.95\linewidth]{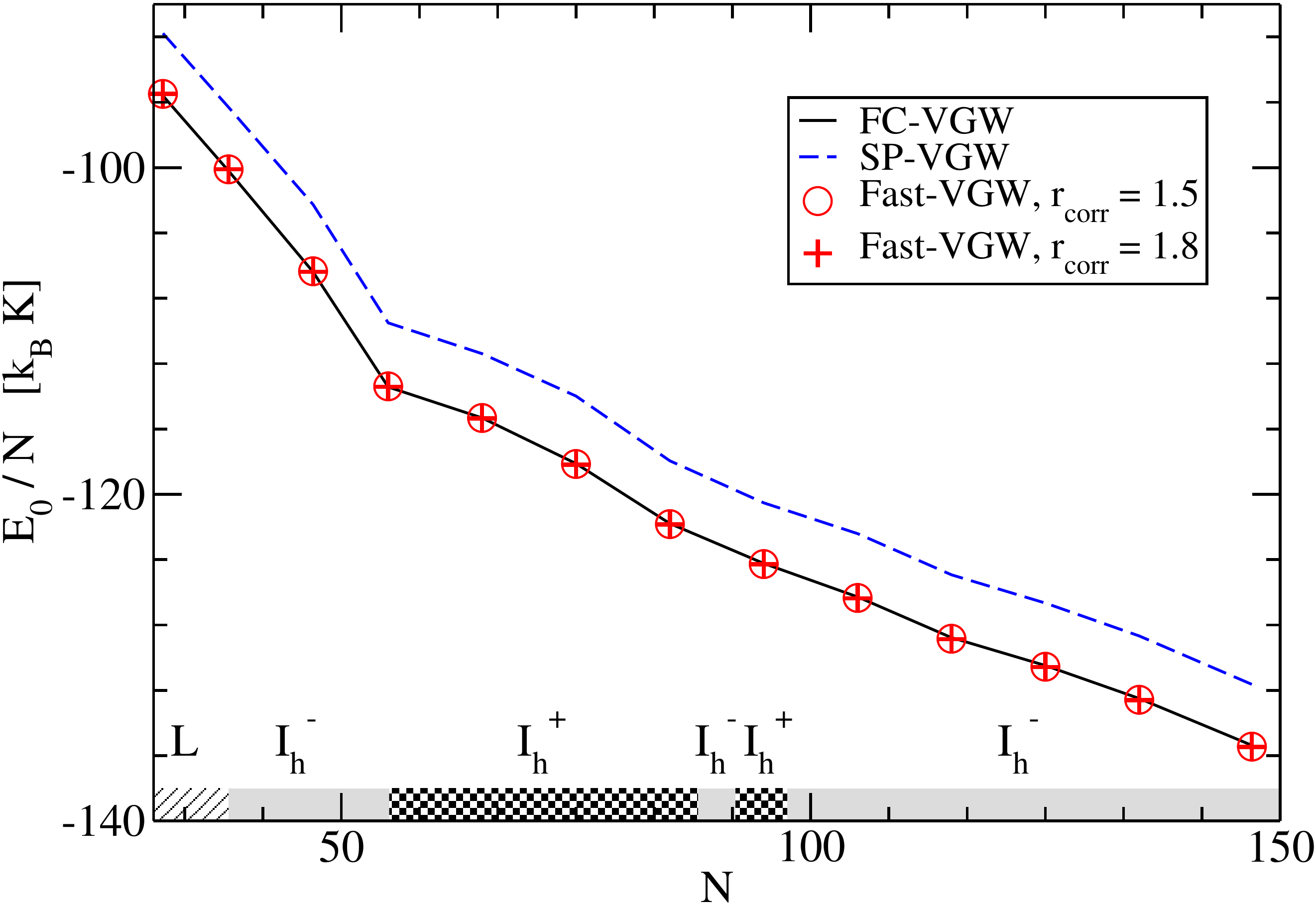}
\end{centering}
\caption{The ground state energy per particle for a sequence of Lennard-Jones clusters as a function of size. The quantum parameter is $\Lambda=0.10$, which approximatively corresponds to Ne. Full line: fully-coupled $3N\times 3N$ Gaussian, dashed blue line: the single particle VGW; open circles: the Fast-VGW with a correlation distance $\rcorr=1.5\sigma$; diamonds: the same, but with $\rcorr=1.8\sigma$. The shading at the bottom indicates the structural motifs of the ground state\cite{deckman2009a}: liquid (L), Mackay icosahedron (I$_h^-$), anti-Mackay icosahedron (I$_h^+$).}\label{qp0.1}
\end{figure}

With increasing quantum parameter, particle correlations gain significance and the off-diagonal blocks of the $G$ matrix should not be neglected. We will thus include the off-diagonal blocks, but only for pairs of particles, less then a certain distance apart,  $x_{ij}<\rcorr$, which makes $G$ sparse but preserves its positive-definiteness. Such an approach is also motivated by Fig. \ref{G147}, which shows the $G$ matrix for a Ne$_{147}$ cluster at $T=3.56 K$ ($T=0.1\epsilon$) within the fully coupled Gaussian (FC-VGW) approach (note the logarithmic color scale). Most  elements of the $G$-matrix are seven orders of magnitude smaller then the dominant ones and can therefore be neglected.

A first example of the sparse $G$  approach (Fast-VGW) is given in Fig. \ref{qp0.1}, which shows the ground state energy per particle $E_0/N$ for a sequence of (LJ)$_N$ clusters as a function of size. The quantum parameter is $\Lambda=0.1$, corresponding roughly to neon. The ground state energy was computed from Eq.~\eqref{gse} after propagating Eqs.~(\ref{vgw:qp}-\ref{vgw:gammap}) to $\beta=100$ (the Gaussian parameters become stationary at $\beta\sim 50$ already). The energies obtained with fast-VGW (symbols) are indistinguishable from that using the full $3N\times 3N$ $G$ matrix (solid black line).
Note that the LJ interaction is typically cut off at radius $\rcutoff=2.75\sigma$, which is much larger than $\rcorr=1.5\sigma$ (open circles).  For $N=147$, the latter value results in a $G$ matrix with 7.2\% non-zero elements, i.e. the sparsity of $G$ can already make a noticeable difference.  Although the single-Gaussian result (the dashed blue line) is displaced, but  it runs parallel to the FC-VGW result and is able to correctly characterize different configurations according to their energies. The shaded areas indicate the structural motif of the ground state according to the $n-\Lambda$ phase diagram reported in Ref. \cite{deckman2009a}.

\begin{figure}[htbp]
\begin{centering}
\includegraphics[width=.95\linewidth]{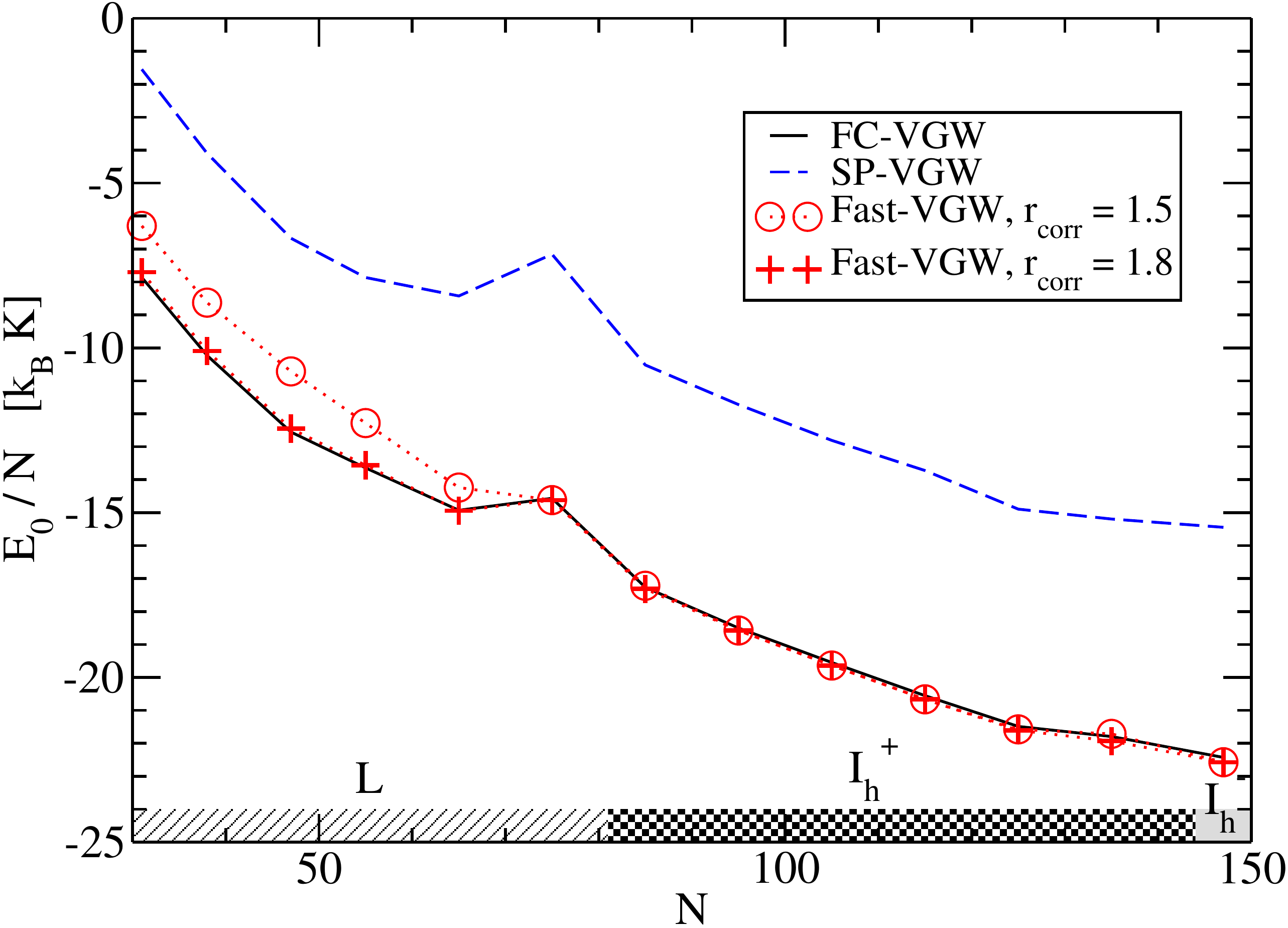}
\end{centering}
\caption{Same as Fig. \ref{qp0.1} but for $\Lambda=0.30$, corresponding to \emph{para}-hydrogen.}\label{qp0.3}
\end{figure}

Fig. \ref{qp0.3} shows the same analysis, but for a larger value of the quantum parameter, $\Lambda=0.3$, which corresponds to \emph{para}-H$_2$. The displacement of the SP-VGW result becomes irregular. The Fast-VGW using $\rcorr=1.5\sigma$ is more accurate, but there is a discrepancy for smaller clusters. As indicated by the shading, the ground state structures for this size range have liquid-like character, meaning that the pair distribution function does not immediately vanish to zero after the nearest neighbor peak. Increasing the correlation cut-off to $\rcorr=1.8\sigma$ removes the discrepancy. The increased correlation cut-off did not change the results for the lower $\Lambda$ in Fig. \ref{qp0.1}.

\begin{figure}[htbp]
\begin{centering}
\includegraphics[width=.95\linewidth]{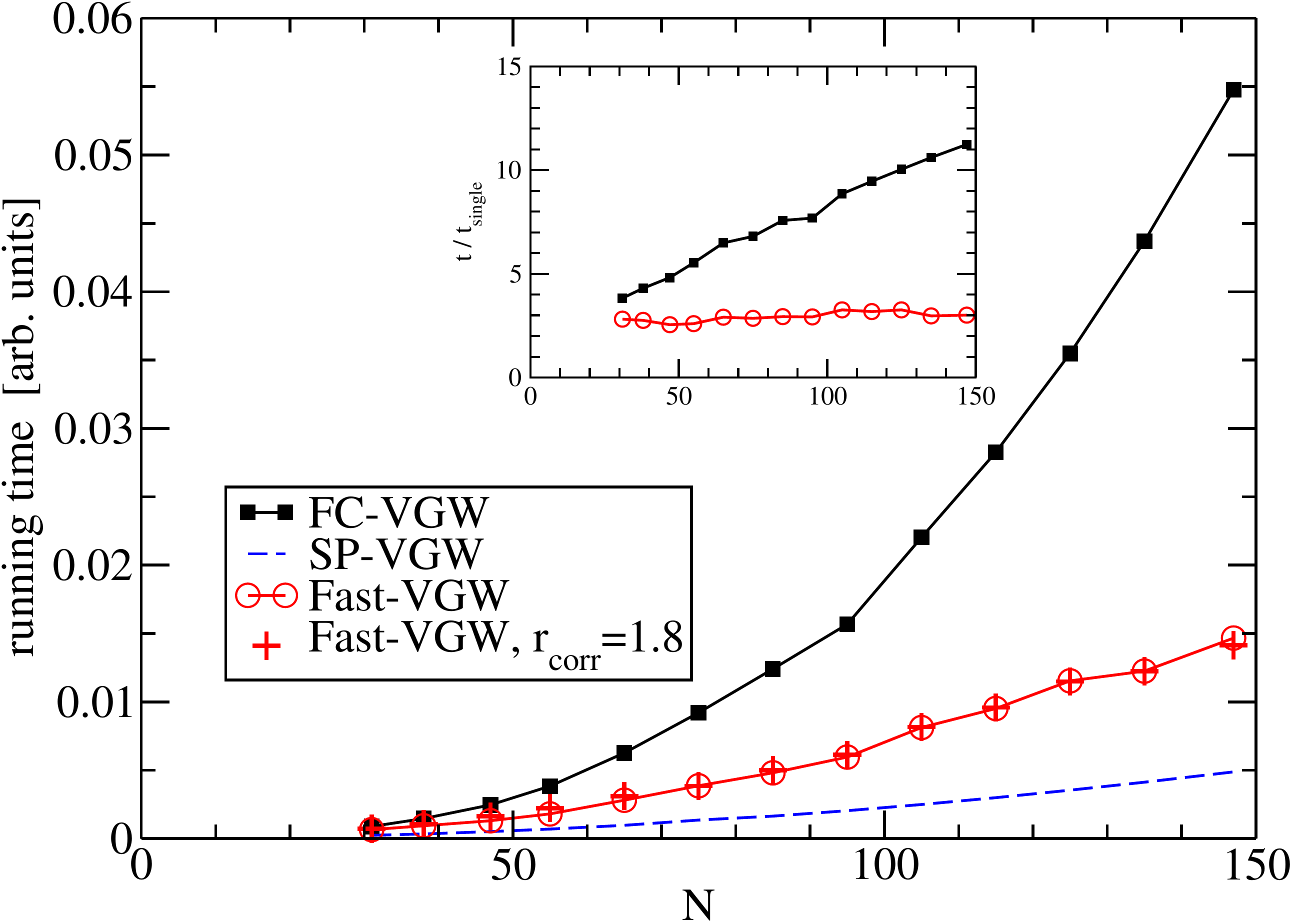}
\end{centering}
\caption{Average time (in arbitrary units) needed to evaluate the right hand side of Eqs.~(\ref{vgw:qp}-\ref{vgw:gammap}). Black squares: fully-coupled Gaussians; dashed blue: single-particle Gaussians; open red circles: Fast-VGW with correlation distance $\rcorr=1.5\sigma$; crosses: Fast-VGW with $\rcorr=1.8\sigma$. The inset compares the running times of the FC-VGW and Fast-VGW implementations normalized by the running time of the SP-VGW one, underscoring the $\mathcal{O}(N^2)$ running time of the Fast-VGW implementation.}\label{rtime}
\end{figure}

Fig. \ref{rtime} compares the computational cost of the three methods presented here: the FC-VGW (full $G$-matrix), SP-VGW (block-diagonal $G$-matrix) and  Fast-VGW (sparse $G$-matrix). The inset shows the running times for the FC-VGW and the Fast-VGW implementations normalized by the running time of the SP-VGW implementation. The latter is obviously the fastest, but the Fast-VGW is slower by a constant factor and falls in the same class of $\mathcal{O}(N^2)$ scaling. No cut-off radius was used in estimating the Hessian $\langle \nabla\nabla\transpose U \rangle$ in Eq.~\eqref{vgw:gp}, which is $\mathcal{O}(N^2)$ and by far the dominant part of the SP-VGW. The linear increase in the inset confirms the $\mathcal{O}(N^3)$ cost of the full matrix approach. One should stress that the Hessian $\langle \nabla\nabla\transpose U \rangle$ is still a dense matrix. Applying the usual LJ cutoff $\rcutoff=2.75\sigma$ or an even larger, more ``generous'' one,  would make the Hessian sparse too, which would further improve the performance. The additional cutoff was not applied here to keep the implementation as simple as possible. One should also note that the increased correlation cut-off did not increase the running time significantly.

\section*{Size-induced icosahedral-decahedral transition in large neon clusters.}

The Fast-VGW is so far able to provide the same accuracy as the fully coupled VGW at a fraction of the computational cost.  Following Ref. \cite{doye2002} we applied this new powerful tool to a number of Ne$_N$ clusters with up to N=6500 atoms to study the competition between the structural motifs of the ground state structure.
The ratio of surface atoms decreases with increasing cluster size and therefore, a switch from the predominantly icosahedral motif of small clusters to motifs that optimize the bulk energy over the surface energy is expected. Ref. \cite{doye2002} has estimated the crossover sizes within the Harmonic-Superposition Method (HSM) for both classical and quantum clusters. We expect to improve the quantum mechanical result of the HSM because the VGW accounts inherently for anharmonic contributions and relaxes, at the same time, the structure of the cluster as the quantum particles become ``larger'' (more delocalized). Since the VGW method is variational and is exact for a harmonic potential, it should always provide better upper-bound energy estimates than the HSM method.

A detailed structural analysis of large clusters as a function of their size is practically unfeasible even in the classical case, as the number of local minima increases exponentially with size, and is already enormous for several tens of atoms; Global optimization methods have not broken the $N=1000$ barrier yet \cite{doye2002,xiang2004,xiang2004a,wu2009,ccd2011}. Moreover, except for a better accuracy of the crossover size, a detailed picture would not contribute much to understanding the physics of the crossover.
Thus, as in Ref. \cite{doye2002}, we have also settled for a coarse-grained picture, following the sequence of lowest energy configurations of Mackay icosahedra and Marks decahedra. The icosahedral motif actually favors truncated structures for $N\geq 923$, corresponding to the ``magic'' number sequence minus the 12 vertices. Consequently, We have studied $N=911$, 1403, 2045, 2857, 3859, 5071 and 6513. Similarly, for large sizes the Marks decahedra favor truncated structures too, so we have studied $N=1103$, 1660, 2377, 3274 and 4371. We evaluated the energies according to Eq.~\eqref{gse} after propagating Eqs.~(\ref{vgw:qp}-\ref{vgw:gammap}) to $\beta=100$. 
Within each structural type, 
the obtained ground state energy values are then used to interpolate the energy as a function 
of $N$ according to:
\begin{equation}
E(N) = a_E N + b_E N^{2/3} + c_E N^{1/3} + d_E
\label{energy:fit}
\end{equation}
where $a_E$ denotes the bulk and  $b_E$ the surface contribution, respectively.

\begin{figure}[htbp]
\begin{centering}
\includegraphics[width=.95\linewidth]{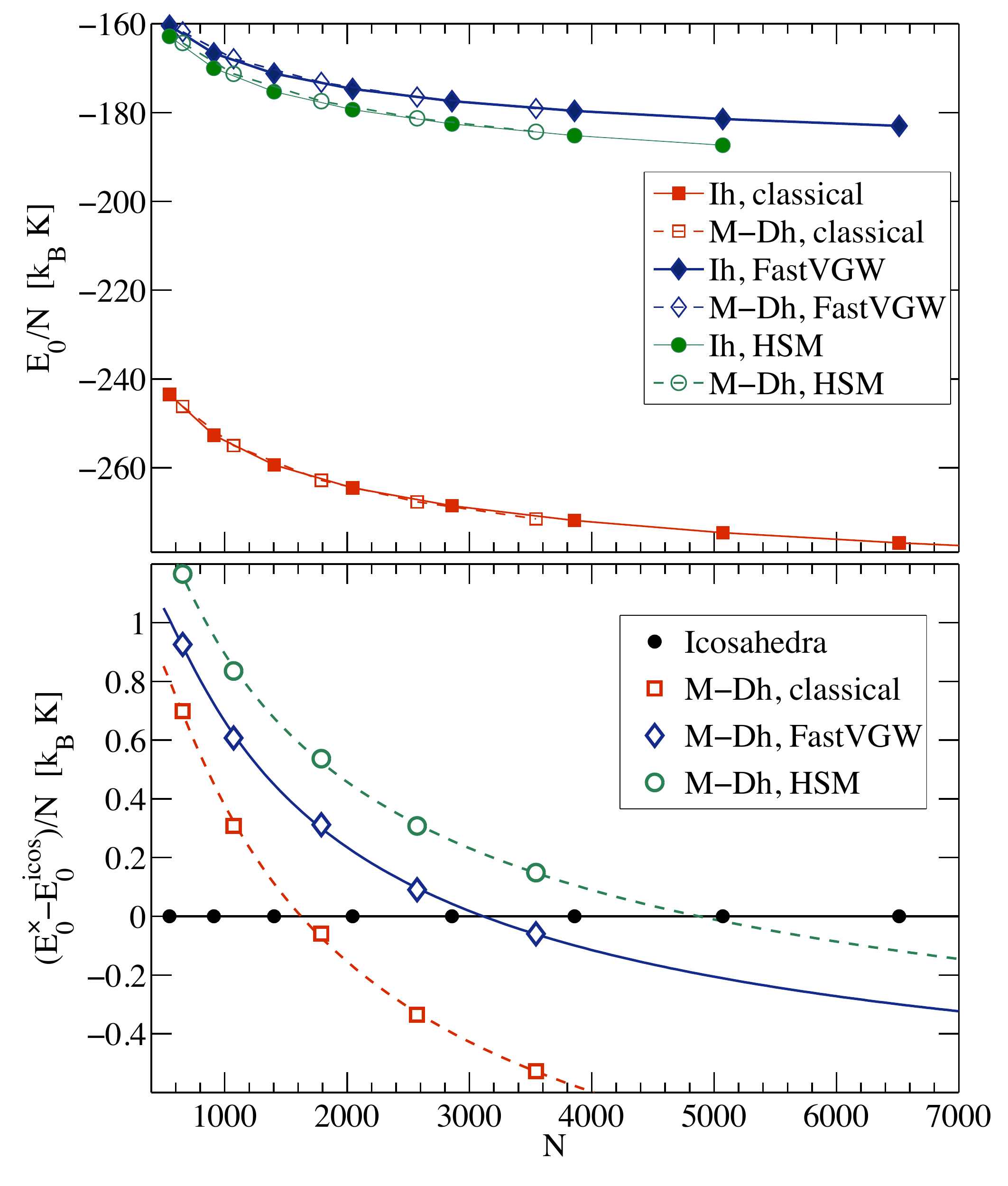}
\end{centering}
\caption{The ground-state energy per particle $E_0/N$ of Mackay icosahedra versus Marks decahedra estimated by the Fast-VGW and the HSM approach. Filled symbols: icosahedra (I$_h$); empty symbols: Marks decahedra (M-D$_h$). Upper panel: absolute energies. Lower panel: the energy of the decahedral motifs relative the icosahedral ones; in this panel the lines show interpolations according to Eq. \eqref{energy:fit}. }\label{icos_vs_md}
\end{figure}

\begin{figure}[htbp]
\includegraphics[width=\linewidth]{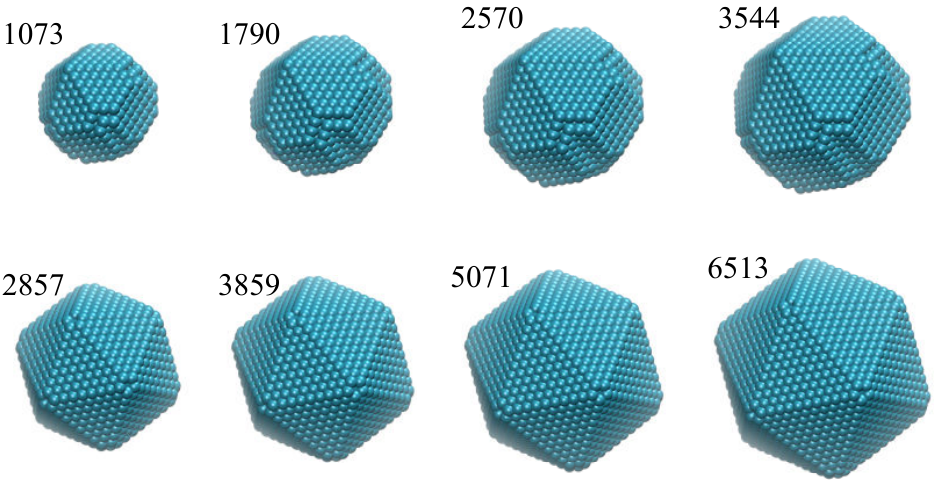}
\caption{The four largest structures investigated in this work for each of the
structural motifs. Upper row: truncated Marks-decahedra; lower row: truncated
icosahedra. From Ref.\cite{doye2002}.}\label{structures}
\end{figure}

Fig. \ref{icos_vs_md} shows the computed ground state energies for the
icosahedral and the decahedral motifs of (Ne)$_N$ clusters estimated by means
of the Fast-VGW and of the Harmonic-Superposition Method (HSM). The VGW
energies are systematically higher (HSM is not variational) and provide a
correct upper bound to the ground-state energy. Being exact in the harmonic
limit, the VGW would have yielded the HSM result, had the anharmonic
contribution been insignificant. One should also note that the VGW  relaxes
the cluster structure, which typically increases in size slightly due to the
quantum delocalization. The classical result is also included to quantify the
zero-point energy effect. The largest structures investigated here are
illustrated in Fig.  \ref{structures}.

The crossover is almost imperceptible on an absolute energy scale. The lower panel of Fig. \ref{icos_vs_md} shows the relative energy of the decahedral sequence with respect to the icosahedral sequence. The relative energy of the icosahedral sequence to itself is obviously zero. Using the interpolation \eqref{energy:fit}, we estimate
the crossover at $N\sim 3100$. This is significantly smaller than the quantum HSM result $N=4640$ and underscores  the significance of the anharmonic effects. Fig. \ref{icos_vs_md} also shows the energy of the classical decahedral motif relative to the classical icosahedral motif. The crossover occurs much earlier, at $N\sim 1690$.  As shown in ref. \cite{doye2002}, the vibrational frequencies are lower for icosahedral structures than for decahedral ones. With increasing quantum parameter $\Lambda$, the zero-point energy of the vibrational modes stabilizes the icosahedra with respect to decahedra and pushes the crossover to larger $N$. Qualitatively identical behavior has also been observed with the unoptimized (non-truncated) structures  (not shown here). The classical crossover is $N\sim 2200$, the VGW predicts $N\sim5400$ and the HSM estimate is $N\sim 10000$.

\section*{Conclusions.}

We have introduced a rather straightforward, but practically significant improvement of the VGW method based on discarding long-range correlations in the representation of the (otherwise fully-coupled) Gaussian wavepacket. The computational advantage is tremendous, reducing the overall cost from $\mathcal{O}(N^3)$ to $\mathcal{O}(N^2)$ with no visible accuracy tradeoff. The new method allowed us to study neon clusters with up to 6500 atoms. Consequently, we were able to directly investigate the competition between the Mackay-icosahedral and Marks-decahedral structural motifs when the cluster size is varied.  Additional sparsity in the Hessian could also be exploited, reducing the numerical scaling even further, to $\mathcal{O}(N)$. Our implementation can easily take advantage of existent massively parallel sparse matrix packages, and as such can be applied to systems with tens of thousands of particles.

\section*{Acknowledgements}
We would like to thank Florent Calvo and Jon Doye for discussions and for providing us with the optimized structures of LJ clusters used in this work. We also thank Aydin Bulu\c{c} for guidance with software packages for sparse matrices.


\begin{thebibliography}{22}
\expandafter\ifx\csname natexlab\endcsname\relax\def\natexlab#1{#1}\fi
\expandafter\ifx\csname bibnamefont\endcsname\relax
  \def\bibnamefont#1{#1}\fi
\expandafter\ifx\csname bibfnamefont\endcsname\relax
  \def\bibfnamefont#1{#1}\fi
\expandafter\ifx\csname citenamefont\endcsname\relax
  \def\citenamefont#1{#1}\fi
\expandafter\ifx\csname url\endcsname\relax
  \def\url#1{\texttt{#1}}\fi
\expandafter\ifx\csname urlprefix\endcsname\relax\def\urlprefix{URL }\fi
\providecommand{\bibinfo}[2]{#2}
\providecommand{\eprint}[2][]{\url{#2}}

\bibitem[{\citenamefont{Frantsuzov et~al.}(2003)\citenamefont{Frantsuzov,
  Neumaier, and Mandelshtam}}]{frantsuzov2003}
\bibinfo{author}{\bibfnamefont{P.}~\bibnamefont{Frantsuzov}},
  \bibinfo{author}{\bibfnamefont{A.}~\bibnamefont{Neumaier}}, \bibnamefont{and}
  \bibinfo{author}{\bibfnamefont{V.~A.} \bibnamefont{Mandelshtam}},
  \bibinfo{journal}{Chemical Physics Letters} \textbf{\bibinfo{volume}{381}},
  \bibinfo{pages}{117 } (\bibinfo{year}{2003}), ISSN \bibinfo{issn}{0009-2614},
  \urlprefix\url{http://www.sciencedirect.com/science/article/B6TFN-49SFJD2-4/2/16b54a93ffe5dea17caddeb61447f6aa}.

\bibitem[{\citenamefont{Frantsuzov and Mandelshtam}(2004)}]{frantsuzov2004}
\bibinfo{author}{\bibfnamefont{P.~A.} \bibnamefont{Frantsuzov}}
  \bibnamefont{and} \bibinfo{author}{\bibfnamefont{V.~A.}
  \bibnamefont{Mandelshtam}}, \bibinfo{journal}{J. Comp. Phys.}
  \textbf{\bibinfo{volume}{121}}, \bibinfo{pages}{9247} (\bibinfo{year}{2004}),
  \urlprefix\url{http://link.aip.org/link/?JCP/121/9247/1}.

\bibitem[{\citenamefont{Frantsuzov et~al.}(2006)\citenamefont{Frantsuzov,
  Meluzzi, and Mandelshtam}}]{frantsuzov2006}
\bibinfo{author}{\bibfnamefont{P.~A.} \bibnamefont{Frantsuzov}},
  \bibinfo{author}{\bibfnamefont{D.}~\bibnamefont{Meluzzi}}, \bibnamefont{and}
  \bibinfo{author}{\bibfnamefont{V.~A.} \bibnamefont{Mandelshtam}},
  \bibinfo{journal}{Phys. Rev. Lett.} \textbf{\bibinfo{volume}{96}},
  \bibinfo{pages}{113401} (\bibinfo{year}{2006}).

\bibitem[{\citenamefont{Deckman et~al.}(2008)\citenamefont{Deckman, Frantsuzov,
  and Mandelshtam}}]{deckman2008}
\bibinfo{author}{\bibfnamefont{J.}~\bibnamefont{Deckman}},
  \bibinfo{author}{\bibfnamefont{P.~A.} \bibnamefont{Frantsuzov}},
  \bibnamefont{and} \bibinfo{author}{\bibfnamefont{V.~A.}
  \bibnamefont{Mandelshtam}}, \bibinfo{journal}{Phys. Rev. E}
  \textbf{\bibinfo{volume}{77}}, \bibinfo{pages}{052102}
  (\bibinfo{year}{2008}).

\bibitem[{\citenamefont{Frantsuzov and Mandelshtam}(2008)}]{frantsuzov2008}
\bibinfo{author}{\bibfnamefont{P.~A.} \bibnamefont{Frantsuzov}}
  \bibnamefont{and} \bibinfo{author}{\bibfnamefont{V.~A.}
  \bibnamefont{Mandelshtam}}, \bibinfo{journal}{J. Comp. Phys.}
  \textbf{\bibinfo{volume}{128}}, \bibinfo{eid}{094304}
  (pages~\bibinfo{numpages}{7}) (\bibinfo{year}{2008}),
  \urlprefix\url{http://link.aip.org/link/?JCP/128/094304/1}.

\bibitem[{\citenamefont{Deckman and
  Mandelshtam}(2009{\natexlab{a}})}]{deckman2009}
\bibinfo{author}{\bibfnamefont{J.}~\bibnamefont{Deckman}} \bibnamefont{and}
  \bibinfo{author}{\bibfnamefont{V.~A.} \bibnamefont{Mandelshtam}},
  \bibinfo{journal}{Phys. Rev. E} \textbf{\bibinfo{volume}{79}},
  \bibinfo{pages}{022101} (\bibinfo{year}{2009}{\natexlab{a}}).

\bibitem[{\citenamefont{Deckman and
  Mandelshtam}(2009{\natexlab{b}})}]{deckman2009a}
\bibinfo{author}{\bibfnamefont{J.}~\bibnamefont{Deckman}} \bibnamefont{and}
  \bibinfo{author}{\bibfnamefont{V.~A.} \bibnamefont{Mandelshtam}},
  \bibinfo{journal}{J. Phys. Chem. A} \textbf{\bibinfo{volume}{113}},
  \bibinfo{pages}{7394} (\bibinfo{year}{2009}{\natexlab{b}}),
  \urlprefix\url{http://dx.doi.org/10.1021/jp900095f}.

\bibitem[{\citenamefont{Deckman and Mandelshtam}(2010)}]{deckman2010}
\bibinfo{author}{\bibfnamefont{J.}~\bibnamefont{Deckman}} \bibnamefont{and}
  \bibinfo{author}{\bibfnamefont{V.~A.} \bibnamefont{Mandelshtam}},
  \bibinfo{journal}{The Journal of Physical Chemistry A}
  \textbf{\bibinfo{volume}{114}}, \bibinfo{pages}{9820} (\bibinfo{year}{2010}),
  \eprint{http://pubs.acs.org/doi/pdf/10.1021/jp102898b},
  \urlprefix\url{http://pubs.acs.org/doi/abs/10.1021/jp102898b}.

\bibitem[{\citenamefont{Predescu et~al.}(2005)\citenamefont{Predescu,
  Frantsuzov, and Mandelshtam}}]{predescu2005}
\bibinfo{author}{\bibfnamefont{C.}~\bibnamefont{Predescu}},
  \bibinfo{author}{\bibfnamefont{P.~A.} \bibnamefont{Frantsuzov}},
  \bibnamefont{and} \bibinfo{author}{\bibfnamefont{V.~A.}
  \bibnamefont{Mandelshtam}}, \bibinfo{journal}{J. Comp. Phys.}
  \textbf{\bibinfo{volume}{122}}, \bibinfo{eid}{154305}
  (pages~\bibinfo{numpages}{12}) (\bibinfo{year}{2005}),
  \urlprefix\url{http://link.aip.org/link/?JCP/122/154305/1}.

\bibitem[{\citenamefont{Georgescu and Mandelshtam}(2010)}]{georgescu2010}
\bibinfo{author}{\bibfnamefont{I.}~\bibnamefont{Georgescu}} \bibnamefont{and}
  \bibinfo{author}{\bibfnamefont{V.~A.} \bibnamefont{Mandelshtam}},
  \bibinfo{journal}{Phys. Rev. B} \textbf{\bibinfo{volume}{82}},
  \bibinfo{pages}{094305} (\bibinfo{year}{2010}).

\bibitem[{\citenamefont{Georgescu et~al.}(2011)\citenamefont{Georgescu,
  Deckman, Fredrickson, and Mandelshtam}}]{georgescu2011}
\bibinfo{author}{\bibfnamefont{I.}~\bibnamefont{Georgescu}},
  \bibinfo{author}{\bibfnamefont{J.}~\bibnamefont{Deckman}},
  \bibinfo{author}{\bibfnamefont{L.~J.} \bibnamefont{Fredrickson}},
  \bibnamefont{and} \bibinfo{author}{\bibfnamefont{V.~A.}
  \bibnamefont{Mandelshtam}}, \bibinfo{journal}{The Journal of Chemical
  Physics} \textbf{\bibinfo{volume}{134}}, \bibinfo{eid}{174109}
  (pages~\bibinfo{numpages}{9}) (\bibinfo{year}{2011}),
  \urlprefix\url{http://link.aip.org/link/?JCP/134/174109/1}.

\bibitem[{\citenamefont{Liu and Miller}(2007)}]{liu2007}
\bibinfo{author}{\bibfnamefont{J.}~\bibnamefont{Liu}} \bibnamefont{and}
  \bibinfo{author}{\bibfnamefont{W.~H.} \bibnamefont{Miller}},
  \bibinfo{journal}{The Journal of Chemical Physics}
  \textbf{\bibinfo{volume}{126}}, \bibinfo{eid}{234110}
  (pages~\bibinfo{numpages}{11}) (\bibinfo{year}{2007}),
  \urlprefix\url{http://link.aip.org/link/?JCP/126/234110/1}.

\bibitem[{\citenamefont{Liu and Miller}(2011)}]{liu2011a}
\bibinfo{author}{\bibfnamefont{J.}~\bibnamefont{Liu}} \bibnamefont{and}
  \bibinfo{author}{\bibfnamefont{W.~H.} \bibnamefont{Miller}},
  \bibinfo{journal}{The Journal of Chemical Physics}
  \textbf{\bibinfo{volume}{134}}, \bibinfo{eid}{104102}
  (pages~\bibinfo{numpages}{19}) (\bibinfo{year}{2011}),
  \urlprefix\url{http://link.aip.org/link/?JCP/134/104102/1}.

\bibitem[{\citenamefont{Liu}(2011)}]{liu2011b}
\bibinfo{author}{\bibfnamefont{J.}~\bibnamefont{Liu}}, \bibinfo{journal}{The
  Journal of Chemical Physics} \textbf{\bibinfo{volume}{134}},
  \bibinfo{eid}{194110} (pages~\bibinfo{numpages}{19}) (\bibinfo{year}{2011}),
  \urlprefix\url{http://link.aip.org/link/?JCP/134/194110/1}.

\bibitem[{\citenamefont{Doye and Calvo}(2002)}]{doye2002}
\bibinfo{author}{\bibfnamefont{J.~P.~K.} \bibnamefont{Doye}} \bibnamefont{and}
  \bibinfo{author}{\bibfnamefont{F.}~\bibnamefont{Calvo}},
  \bibinfo{journal}{The Journal of Chemical Physics}
  \textbf{\bibinfo{volume}{116}}, \bibinfo{pages}{8307} (\bibinfo{year}{2002}),
  \urlprefix\url{http://link.aip.org/link/?JCP/116/8307/1}.

\bibitem[{\citenamefont{Silvera and Goldman}(1978)}]{silvera1978}
\bibinfo{author}{\bibfnamefont{I.~F.} \bibnamefont{Silvera}} \bibnamefont{and}
  \bibinfo{author}{\bibfnamefont{V.~V.} \bibnamefont{Goldman}},
  \bibinfo{journal}{J. Comp. Phys.} \textbf{\bibinfo{volume}{69}},
  \bibinfo{pages}{4209} (\bibinfo{year}{1978}),
  \urlprefix\url{http://link.aip.org/link/?JCP/69/4209/1}.

\bibitem[{\citenamefont{Barnes and Hut}(1986)}]{barnes1986}
\bibinfo{author}{\bibfnamefont{J.}~\bibnamefont{Barnes}} \bibnamefont{and}
  \bibinfo{author}{\bibfnamefont{P.}~\bibnamefont{Hut}},
  \bibinfo{journal}{Nature} \textbf{\bibinfo{volume}{324}},
  \bibinfo{pages}{446} (\bibinfo{year}{1986}).

\bibitem[{\citenamefont{Greengard and Rokhlin}(1987)}]{greengard1987}
\bibinfo{author}{\bibfnamefont{L.}~\bibnamefont{Greengard}} \bibnamefont{and}
  \bibinfo{author}{\bibfnamefont{V.}~\bibnamefont{Rokhlin}},
  \bibinfo{journal}{Journal of Computational Physics}
  \textbf{\bibinfo{volume}{73}}, \bibinfo{pages}{325} (\bibinfo{year}{1987}),
  \urlprefix\url{http://www.sciencedirect.com/science/article/B6WHY-4DD1T30-K7/2/2b3def8a3a8d71ff0d1697298ea6d2c8}.

\bibitem[{\citenamefont{Xiang et~al.}(2004{\natexlab{a}})\citenamefont{Xiang,
  Jiang, Cai, and Shao}}]{xiang2004}
\bibinfo{author}{\bibfnamefont{Y.}~\bibnamefont{Xiang}},
  \bibinfo{author}{\bibfnamefont{H.}~\bibnamefont{Jiang}},
  \bibinfo{author}{\bibfnamefont{W.}~\bibnamefont{Cai}}, \bibnamefont{and}
  \bibinfo{author}{\bibfnamefont{X.}~\bibnamefont{Shao}}, \bibinfo{journal}{J.
  Phys. Chem. A} \textbf{\bibinfo{volume}{108}}, \bibinfo{pages}{3586}
  (\bibinfo{year}{2004}{\natexlab{a}}), ISSN \bibinfo{issn}{1089-5639},
  \urlprefix\url{http://pubs3.acs.org/acs/journals/doilookup?in_doi=10.1021/jp037780t}.

\bibitem[{\citenamefont{Xiang et~al.}(2004{\natexlab{b}})\citenamefont{Xiang,
  Cheng, Cai, and Shao}}]{xiang2004a}
\bibinfo{author}{\bibnamefont{Xiang}}, \bibinfo{author}{\bibnamefont{Cheng}},
  \bibinfo{author}{\bibnamefont{Cai}}, \bibnamefont{and}
  \bibinfo{author}{\bibnamefont{Shao}}, \bibinfo{journal}{The Journal of
  Physical Chemistry A} \textbf{\bibinfo{volume}{108}}, \bibinfo{pages}{9516}
  (\bibinfo{year}{2004}{\natexlab{b}}),
  \eprint{http://pubs.acs.org/doi/pdf/10.1021/jp047807o},
  \urlprefix\url{http://pubs.acs.org/doi/abs/10.1021/jp047807o}.

\bibitem[{\citenamefont{Wu et~al.}(2009)\citenamefont{Wu, Cai, and
  Shao}}]{wu2009}
\bibinfo{author}{\bibfnamefont{X.}~\bibnamefont{Wu}},
  \bibinfo{author}{\bibfnamefont{W.}~\bibnamefont{Cai}}, \bibnamefont{and}
  \bibinfo{author}{\bibfnamefont{X.}~\bibnamefont{Shao}},
  \bibinfo{journal}{Chemical Physics} \textbf{\bibinfo{volume}{363}},
  \bibinfo{pages}{72 } (\bibinfo{year}{2009}), ISSN \bibinfo{issn}{0301-0104},
  \urlprefix\url{http://www.sciencedirect.com/science/article/pii/S0301010409002407}.

\bibitem[{\citenamefont{Wales et~al.}()\citenamefont{Wales, Doye, Dullweber,
  Hodges, Naumkin, Calvo, Hern{\'a}ndez-Rojas, and Middleton}}]{ccd2011}
\bibinfo{author}{\bibfnamefont{D.~J.} \bibnamefont{Wales}},
  \bibinfo{author}{\bibfnamefont{J.~P.~K.} \bibnamefont{Doye}},
  \bibinfo{author}{\bibfnamefont{A.}~\bibnamefont{Dullweber}},
  \bibinfo{author}{\bibfnamefont{M.~P.} \bibnamefont{Hodges}},
  \bibinfo{author}{\bibfnamefont{F.~Y.} \bibnamefont{Naumkin}},
  \bibinfo{author}{\bibfnamefont{F.}~\bibnamefont{Calvo}},
  \bibinfo{author}{\bibfnamefont{J.}~\bibnamefont{Hern{\'a}ndez-Rojas}},
  \bibnamefont{and} \bibinfo{author}{\bibfnamefont{T.~F.}
  \bibnamefont{Middleton}}, \emph{\bibinfo{title}{The {C}ambridge {C}luster
  {D}atabase}}, \urlprefix\url{http://www-wales.ch.cam.ac.uk/CCD.html}.

\end{thebibliography}

\end{document}